\documentclass[twocolumn,superscriptaddress,showpacs,preprintnumbers,amsmath,amssymb]{revtex4}

\usepackage{latexsym}% Basic package(s)
\usepackage{graphicx}% Include figure files
\usepackage{dcolumn}% Align table columns on decimal point
\newcommand{\bra}[1]{\langle #1 |}
\newcommand{\brav}[1]{\bra{\mathbf{V}_{#1}}}
\newcommand{\bravp}[1]{\bra{\mathbf{V}^{\prime}_{#1}}}
\newcommand{\ket}[1]{| #1 \rangle}
\newcommand{\ketv}[1]{\ket{\mathbf{V}_{#1}}}
\newcommand{\ketvp}[1]{\ket{\mathbf{V}^{\prime}_{#1}}}

\newcommand{\im}{\dot{\iota}\,}
\newcommand{\var}[1]{\vartriangle \hspace{-0.1cm} #1}
\begin{document}

\title{Multipartite W states for chains of atoms conveyed
       through an optical cavity}

\author{D. Gon\c{t}a}
\email{gonta@physi.uni-heidelberg.de}
\affiliation{Max-Planck-Institut f\"{u}r Kernphysik, P.O.~Box
             103980, D-69029 Heidelberg, Germany}

\author{S. Fritzsche}
\email{s.fritzsche@gsi.de} \affiliation{Department of Physical
Sciences, P.O.~Box 3000,
             Fin-90014 University of Oulu, Finland}%
\affiliation{GSI Helmholtzzentrum f\"{u}r Schwerionenforschung,
             D-64291 Darmstadt, Germany}%

\date{\today}

\begin{abstract}
We propose and work out a scheme to generate the entangled W states
for a chain of $N$ four-level atoms which are transported through an
optical cavity by means of an optical lattice. This scheme is based
on the combined laser-cavity mediated interaction between distant
and equally separated atoms and works in a completely deterministic
way for qubits encoded by two hyperfine levels of the atoms. Only
two parameters, namely the distance between the atoms and the
velocity of the chain, determine the effective interaction among the
atoms and, therefore, the degree of entanglement that is obtained
for the overall chain of $N$ qubits. In particular, we work out the
parameter regions for which the W$_N$ states are generated most
reliably for chains of $N = 2,3,4$ and $5$ atoms. In addition, we
analyze the sensitivity in the formation of entanglement for such
chains of qubits due to uncertainties produced by the oscillations
of atoms in optical lattices.
\end{abstract}


\maketitle

\section{Introduction}

During the last decades, quantum entanglement has been found
essential not only in studying the non-classical behavior of
composite systems but also as important resource for the engineering
and processing of quantum information. Nowadays, numerous
applications are known that greatly benefit from having entangled
quantum states available, such as super-dense coding \cite{bew},
quantum cryptography \cite{eke}, or Grover's quantum search
algorithm \cite{gro} to name just a few. Despite of the recent
progress in dealing with composite quantum systems, however, the
controlled manipulation of these system and their interaction with
the environment remains still a great challenge. Beside of various
other implementations of composite quantum systems for quantum
control and applications in quantum information, a excellent control
in generating entangled states has been achieved recently with
neutral atoms that are coupled to a high-finesse optical cavity
\cite{prl102, prl103}.

In practice, there are two typical ways to encode a single qubit
into the level structure of an atom: Apart from (i) selecting two
levels separated by an optical transition frequency
(\textit{optical} qubit), one may also utilize two hyperfine levels
of--typically--the atomic ground state, sometimes referred to as
\textit{hyperfine} qubit. In contrast to the optical qubits, the use
of hyperfine qubits has the advantage of long coherence times
($\sim$ 1 s) and, moreover, such qubits are known to be more robust
with regard to external perturbations or stray fields. Finally, a
number of microwave techniques have been developed during the last
decades that allow to initialize, manipulate, and detect the states
of such qubits \cite{pra75, prl93, prl96}.

\begin{figure}
\begin{center}
\includegraphics[width=0.455\textwidth]{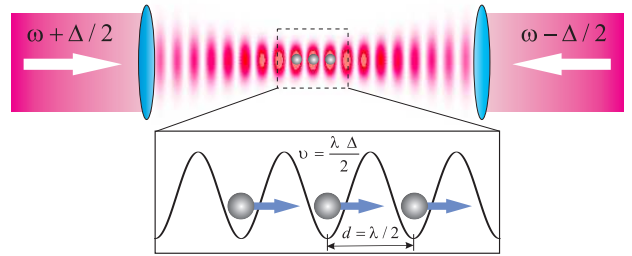} \\
\vspace{0.1cm} \caption{(Color online) Schematic view of atoms in an
optical lattice (conveyor belt). Two focussed and counter-propagating
laser beams with frequencies $\omega + \Delta/2$ and $\omega -
\Delta/2$ give rise to an interference pattern in the field strength
with a series of equidistant potential wells in which the atoms can
be trapped. The distance $d$ between two neighbored wells is given
by (half of) the lattice wavelength $\lambda$, while the velocity of
the belt $\upsilon$ is determined by the detuning $\Delta$ of the
two laser beams.}
\label{fig:1}
\end{center}
\end{figure}

However, the hyperfine qubit(s) cannot couple directly to a cavity
with a resonant mode-frequency in the optical domain. Therefore, a
four-level configuration need to be considered, in which the two
hyperfine levels are supplemented by two electronically excited
levels which are separated by the optical transition frequency and
being compatible with the resonant frequency of the cavity. In order
to realize an effective manipulation of the hyperfine qubit by means
of the optical cavity, its state must be mapped coherently upon the
electronically excited states and back to the hyperfine levels, once
all the desired atom-cavity interactions have been performed.
Indeed, such a indirect coupling between the hyperfine qubit and
optical cavity opens a route for the generation of entanglement and
more complex quantum states between two or more atoms by means of
the cavity-mediated interactions.

Despite of the recent progress to couple one single atom to the
cavity mode, further control of the atomic motion is necessary in
order to manipulate the interaction between cavity and a chain of
atoms. At the same time, an excellent control of the motion of
atomic chains is merely possible by using optical lattices (conveyor
belts) \cite{sc293}, which have recently been utilized in various
setups of cavity QED \cite{prl95, prl98, njp10}. In Fig.~\ref{fig:1}
we displayed a schematic view of such optical lattice in which two
counter-propagating laser beams with parallel linear polarization
produce an interference pattern in the field strength that gives
rise to a series of equidistant potential wells, where neutral atoms
can be trapped. These wells allow to control the position of atoms
with a sub-micrometer precision over millimeter distances due to
their tight confinement along the lattice axis.

The combination of such a lattice with the (optical) cavity QED
setup, however, make it necessary to revise the evolution of the
atom-cavity interaction for a chain of atoms that is
\textit{conveyed} by such a lattice through the cavity. In
particular, one need to analyze how the spacing between the atoms
and velocity of the atomic chain (lattice) will affect the formation
of entangled states between the hyperfine qubits. By this revised
evolution, moreover, the small sample approximation, i.e., when the
spacing between the atoms is considered negligible in comparison to
the cavity waist, should be abandoned and the position-dependent
effects should be taken into account.

\begin{figure}
\begin{center}
\includegraphics[width=0.455\textwidth]{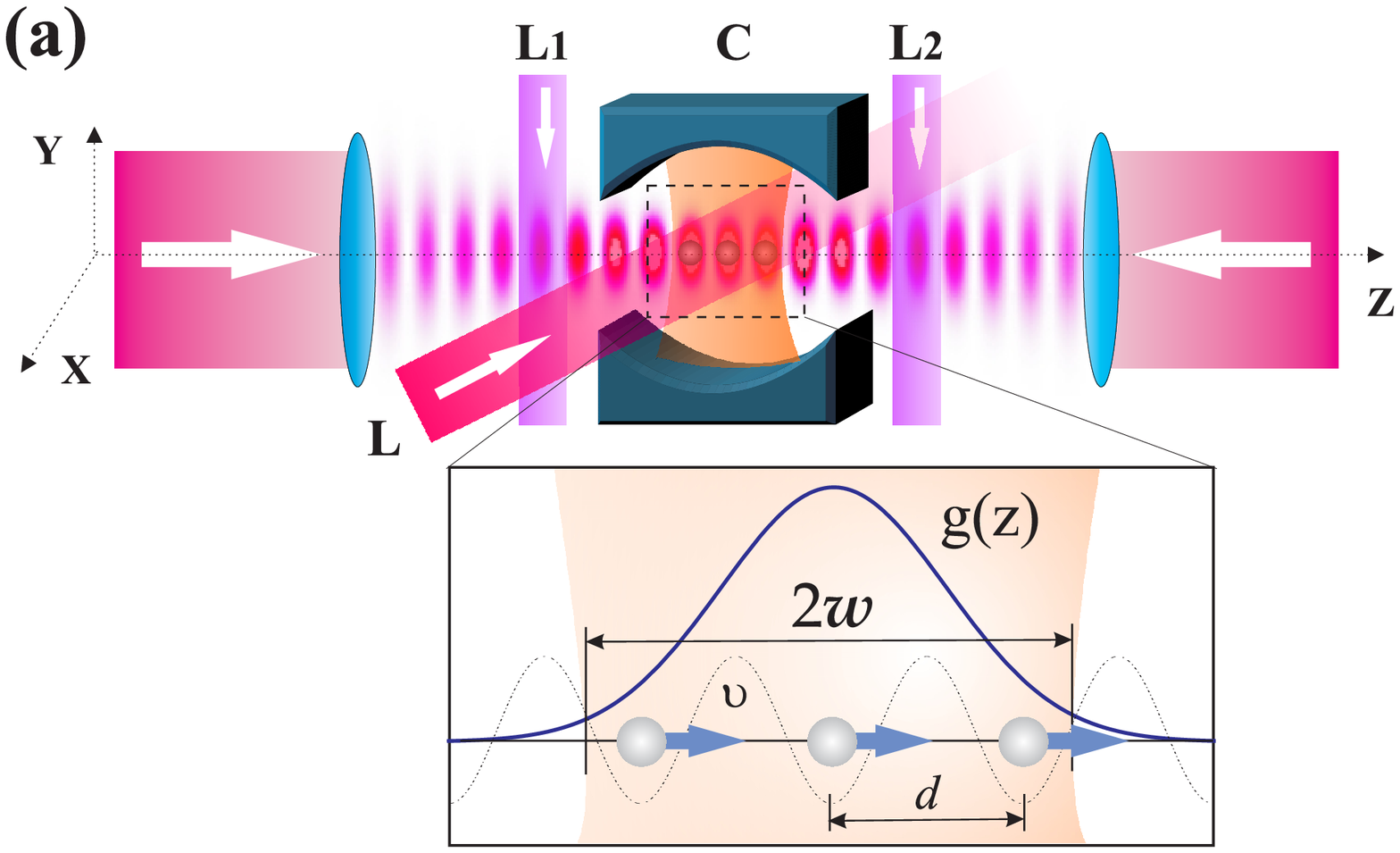} \\
\vspace{0.5cm}
\includegraphics[width=0.43\textwidth]{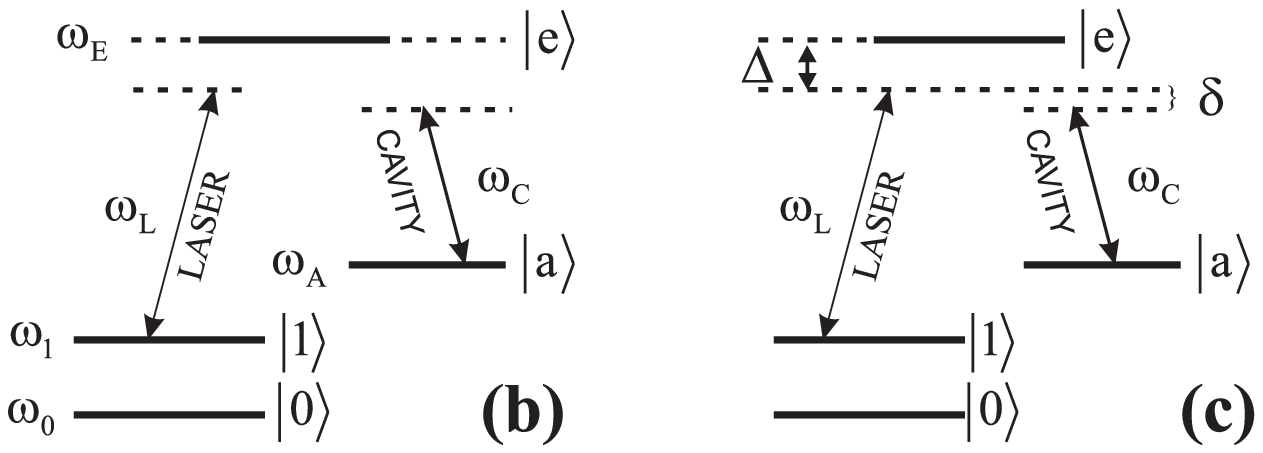} \\
\vspace{0.1cm}
\caption{(Color online) (a) Schematic setup of the experiment. A
chain of $N$ neutral atoms passes through a pair of Raman lasers
$L_1$, an optical cavity $C$ with a laser beam $L$, and a second
pair of Raman lasers $L_2$. The atoms are supposed to move in a
chain along the $z$-axis with a constant velocity $\upsilon$ such
that the chain crosses the cavity at the antinode. Apart from the
cavity waist $w$, that is just one half of the minimum width of the
cavity radiation field, the cavity is also characterized by its
position-dependent coupling strength $g(z)$. (b) The atomic
four-level $\Lambda$-type configuration in the Schr\"{o}dinger
picture and (c) in the interaction picture.}
\label{fig:2}
\end{center}
\end{figure}

For two hyperfine qubits (in a four-level configuration), we have
recently proposed a scheme to generate the maximally entangled state
$\frac{1}{\sqrt{2}} \left( e^{i \phi} \ket{0_1, 1_2} + \ket{1_1,
0_2} \right)$ by means of the combined laser and cavity mediated
interaction \cite{jpb42}. In this reference, we considered the
position-dependent coupling between the atoms and cavity which
allowed us to describe the formation of entanglement between the two
atoms being separated by a macroscopic distance. An effective
interaction between the hyperfine qubits was achieved if both, the
laser and cavity fields, are detuned with regard to the atomic
transition frequencies (see below). In particular, we demonstrated
explicitly how the degree of entanglement depends on the atomic
velocities and the spacing between two atoms.

In the present work, we extend this analysis and propose a scheme to
generate the entangled W state \cite{pra62}
\begin{equation}\label{w-state}
\frac{1}{\sqrt{N}} ( e^{i \phi} \overbrace{\ket{1_1, 0_2, \ldots,
0_N}
                   + \ldots
                   + \ket{0_1, 0_2, \ldots, 1_N}}^{N\rm \;\: terms} )
\end{equation}
between the hyperfine qubits of $N$ (four-level) atoms that are
equally distanced from each other and conveyed through the cavity by
means of an optical lattice. Similarly to Ref.~\cite{jpb42}, this
scheme works in a completely deterministic way and is based on the
position-dependent interaction between distanced atoms which is
mediated by the cavity and laser fields. The two parameters that
control this atom-cavity-laser interaction are (i) the velocity of
the atomic chain along the axis of the lattice and (ii) the distance
between the atoms. For the chains consisting of $N = 2,3,4$ and $5$
atoms, we determine the velocities and distances for which the
initially uncorrelated atoms produce the W$_N$ states most reliably.
Apart from generation of the W states, we discuss also how the
proposed scheme can be implemented most efficiently and analyze how
robust are the entangled states with respect to small oscillations
in the atomic motion as caused by the thermal motion of atoms in the
optical lattice.

The paper is organized as follows. In the next section, we first
outline our scheme to entangle the hyperfine qubits of $N$ initially
uncorrelated atoms. In Section III, we then explain and discuss the
effective Hamiltonian which describes the atomic evolution; we
analyze in particular the parameter (regions) in Subsections
III.A--D for which the states W$_2$, W$_3$, W$_4$, and W$_5$ are
generated most reliably. In Section IV, we later discuss a few
issues related to the implementation of our scheme and how it is
influenced by small oscillations in the motion of the individual
atoms, while a short summary and outlook are given in Section V.

\section{Generation of the W States}

We shall first explain the basic idea of the proposed scheme for
generating multipartite W entangled states within chains of neutral
atoms without going much into details. We assume that the $N$ atoms
are initially in a product state and that they are inserted into an
optical lattice being equally separated by a distance $d$ as
displayed in Fig.~\ref{fig:2}(a). Moreover, the atoms are supposed
to move with a constant velocity $\upsilon$ along the (lattice)
$z$-axis such that their position vectors $\vec{r}_i(t) = \{ 0,0,
z_i^o + \upsilon \, t \}$ cross the cavity at the anti-node and
where $z^o_i$ denote the initial position of the $i-$th atom. As
briefly outlined above [cf.~Fig.~\ref{fig:1}], this velocity
$\upsilon$ and inter-atomic distance $d$ can be controlled
experimentally by adjusting the shift in the frequencies of the two
counter-propagating laser beams and by selecting a proper wavelength
of the optical lattice, respectively \cite{sc293}.

Each of the $N$ identical atoms represents a (hyperfine) qubit in a
$\Lambda$-type configuration as displayed in Fig.~\ref{fig:2}(b), in
which the two hyperfine states $\ket{0}$ and $\ket{1}$ are
supplemented by the electronically excited states $\ket{e}$ and
$\ket{a}$ in such a way, that the transitions $\ket{a} \rightarrow
\ket{1}$ and $\ket{a} \rightarrow \ket{0}$ are (electric-dipole)
forbidden due to the angular momentum and parity selectrion rules.
Initially, the atoms are prepared in the product state
\begin{equation}\label{init-state}
\ket{1_1, 0_2, \ldots, 0_N} \equiv \ket{1_1} \times \ket{0_2} \times
\ldots \times \ket{0_N},
\end{equation}
where the numbering corresponds to the (increasing) coordinates
$z^o_1, z^o_2, \ldots, z^o_N$ of atoms along the $z$-axis. By this
assumption, therefore, the qubits are loaded to the cavity in the
\textit{reverse} order, i.e.\ qubit No.~1 corresponds to the last
atom inside the chain.

Just before each atom enters the cavity, its electronic population
is transferred from the state $\ket{0}$ to the state $\ket{a}$ with
a pair of (slightly) off-resonant laser beams that are coupled to
the atomic transitions $\ket{0} \overset{1}{\rightarrow} \ket{e}
\overset{2}{\rightarrow} \ket{a}$. This population transfer is known
as the two-photon Raman process and can be implemented, for example,
by means of two phase-locked laser diodes \cite{pra75}. Below, we
shall briefly refer to this transfer in the population as the Raman
pulse and shall distinguish between the Raman zones $L_1$ and $L_2$
in front and behind the cavity [see Fig.~\ref{fig:2}(a)]. In this
notation, the Raman pulse $L_2$ is utilized to perform the
population transfer $\ket{a} \overset{2}{\rightarrow} \ket{e}
\overset{1}{\rightarrow} \ket{0}$ back to the hyperfine level.
Inside the cavity, therefore, the product state (\ref{init-state})
becomes $\ket{1_1, a_2, \ldots, a_N}$ since the last atom is
unaffected by the Raman pulse $L_1$.

To explain the mechanism of the cavity-mediated interaction between
the atoms in more detail, let us first consider a chain of two atoms
prepared in the product state $\ket{e_1, a_2}$, and where each atom
is coupled to a detuned optical cavity by the transition $\ket{a}
\leftrightarrow \ket{e}$ [cf.~Fig.~\ref{fig:2}(b)]. In this case,
both atoms interact due to the cavity-stimulated exchange of a
photon: $\ket{e_1, a_2, \bar{0}} \rightarrow \ket{a_1, a_2,\bar{1}}
\rightarrow \ket{a_1, e_2, \bar{0}}$ for a initially empty cavity
$\ket{\bar{0}}$. This interaction sequence contains in its middle
part a virtual state and is independent of the photon number that
was initially in the cavity. By following similar lines, therefore,
the (initial) atomic state $\ket{e_1, a_2, \ldots, a_N}$ of $N$
atoms in the chain evolves according to the sequence
\begin{equation}\label{sequence0}
 \begin{array}{c}
  \ket{e_1, a_2, \ldots, a_N, \bar{0}} \rightarrow
  \ket{a_1, e_2, a_3, \ldots, a_N, \bar{0}}
  \\
  \hspace{2.75cm} \searrow
  \raisebox{-0.15cm}{$\ket{a_1, a_2, e_3, \ldots, a_N, \bar{0}}$}
  \\
  \hspace{2.5cm} \vdots
  \\
  \hspace{3.25cm} \ket{a_1, a_2, a_3, \ldots, e_N, \bar{0}},
  \end{array}
\end{equation}
and where the virtual state $\ket{a_1, \ldots, a_N, \bar{1}}$ has
been omitted for brevity. As seen from the sequence
(\ref{sequence0}), the atoms in the chain interact due to the cavity
stimulated exchange of a single photon between the originally
excited atom and one of the $N-1$ other atoms. This photon exchange,
moreover, requires a rather large detuning between the transition
frequency of the atoms and the resonant frequency of the cavity mode
\begin{equation}\label{condition}
|\,(\omega_E - \omega_A) - \omega_C\,| \gg g(\vec{r}_i), \quad
i = 1, \ldots, N \, ,
\end{equation}
namely such that the cavity remains almost unpopulated in the course
of interaction \cite{prl85}. In the expression above, we have
introduced the position-dependent atom-cavity coupling
\begin{equation}\label{coupling}
g(\vec{r}) = g_o \, \exp \left[ - z^2 / w^2 \right],
\end{equation}
which is caused by the variation of the transversal cavity field
along the atomic trajectories, and where $g_o$ denotes the vacuum
Rabi frequency and $w$ the cavity mode waist, i.e.~one half of the
minimum width of the cavity field [see Fig.~\ref{fig:2}(a)].

Recall that according to our scheme, however, the atoms are loaded
to the cavity in the product state $\ket{1_1, a_2, \ldots, a_N}$ and
hence an intermediate excitation $\ket{1} \rightarrow \ket{e}$ is
first needed to bring the atoms to interaction by means of the
detuned cavity (see above). In order to realize this excitation, the
atomic chain is exposed to a detuned laser beam with frequency
$\omega_L$ that couples --- via a position-independent strength
$\Omega$ to the $\ket{1} \leftrightarrow \ket{e}$ transition as
shown in Fig.~\ref{fig:2}. With the couplings of the atoms to the
both laser and cavity fields, the initial atomic state $\ket{1_1,
a_2, \ldots, a_N}$ evolves according to the sequence of intermediate
atom-cavity states
\begin{widetext}
\begin{equation}\label{sequence}
 \begin{array}{cc}
  & \ket{1_1, a_2, \ldots, a_N, \bar{0}} \rightarrow \ket{e_1, a_2,
  \ldots, a_N, \bar{0}} \rightarrow \ket{a_1, a_2, \ldots, a_N,
  \bar{1}} \rightarrow \ket{a_1, e_2, a_3 \ldots, a_N, \bar{0}} \rightarrow
  \ket{a_1, 1_2, a_3, \ldots, a_N, \bar{0}}
  \\
  & \hspace{9.2cm} \searrow
  \raisebox{-0.15cm}{$\ket{a_1, a_2, e_3, \ldots, a_N, \bar{0}}
  \rightarrow \ket{a_1, a_2, 1_3, \ldots, a_N, \bar{0}}$}
  \\
  & \hspace{10cm} \vdots \hspace{4cm} \vdots
  \\
  & \hspace{9.7cm} \ket{a_1, a_2, a_3, \ldots, e_N, \bar{0}}
  \rightarrow \ket{a_1, a_2, a_3, \ldots, 1_N, \bar{0}}
  \end{array}
\end{equation}
\end{widetext}
into the one of the final states $\ket{a_1, 1_2, a_3, \ldots, a_N},$
$\ldots$, $\ket{a_1, a_2, a_3, \ldots, 1_N}$, which can have the
same probability to occur.

In Refs.~\cite{jpb42, pra67} it was shown that the condition
\begin{equation}\label{condition1}
|(\omega_E - \omega_1) - \omega_L| \gg \Omega
\end{equation}
for the detuning between the atomic transition and laser
frequencies, ensures that the states $\ket{e_1, a_2, \ldots, a_N},$
$\ket{a_1, e_2, \ldots, a_N},$ $ \ldots,$ $\ket{a_1, a_2, \ldots,
e_N}$ remain almost unpopulated. For this reason, the condition
(\ref{condition1}) plays the same role as the condition
(\ref{condition}) for the atom-cavity interaction which makes the
state $\ket{a_1, \ldots, a_N, \bar{1}}$ to be only virtually
populated. In the following, we shall omit these unpopulated
(intermediate) states and express the sequence (\ref{sequence}) in
the short form
\begin{equation}\label{sequence1}
 \begin{array}{c}
  \ket{1_1, a_2, \ldots, a_N, \bar{0}} \rightarrow
  \ket{a_1, 1_2, a_3, \ldots, a_N, \bar{0}}
  \\
  \hspace{2.75cm} \searrow
  \raisebox{-0.15cm}{$\ket{a_1, a_2, 1_3, \ldots, a_N, \bar{0}}$}
  \\
  \hspace{2.5cm} \vdots
  \\
  \hspace{3.25cm} \ket{a_1, a_2, a_3, \ldots, 1_N, \bar{0}}.
  \end{array}
\end{equation}

With the evolution (\ref{sequence1}) of the atomic chain due to the
laser-cavity mediated interaction, the entangled W state
\begin{equation}\label{w-state1}
\frac{1}{\sqrt{N}} ( e^{i \phi} \overbrace{ \ket{1_1, a_2, \ldots,
                                a_N} + \ldots
                                + \ket{a_1, a_2, \ldots, 1_N}}^{N
                     \rm \;\: terms})
\end{equation}
can be, in principle, generated by adjusting properly the atomic
velocity $\upsilon$ and the inter-atomic distance $d$ for a given
set of fixed cavity and laser parameters. In the last step of our
scheme, the electronic population of each atom that leaves the
cavity is (coherently) transferred back from the state $\ket{a}$ to
the hyperfine state $\ket{0}$ in order to protect the atom against
the spontaneous decay. This back transfer is achieved by applying
the Raman pulses $L_2$ behind the cavity and it produces the state
(\ref{w-state}) from (\ref{w-state1}), once all the atoms have left
the cavity. By utilizing the proposed scheme, therefore, a
multipartite W state can be generated starting from the initial
product state (\ref{init-state}) that is associated with the chain
of $N$ atoms which are conveyed through the cavity.

In the next section, we shall analyze in details of how the
laser-cavity mediated evolution (\ref{sequence1}) depends on the
velocity $\upsilon$ and the inter-atomic distance $d$ when the
atomic chain is conveyed through the cavity. In order to take into
account these two parameters, we shall consider the
position-dependent coupling (\ref{coupling}), which gives rise to
the time-dependent coupling between the $i$-th atom and the cavity
\begin{equation}\label{coupling1}
g_i(t) = g_o \, \exp \left[ - (z_i^o + \upsilon \, t)^2 / w^2
\right], \quad i = 1, \ldots, N  \, ,
\end{equation}
where $z^o_i$ and $\upsilon$ denote the initial position of the
$i-$th atom and its velocity along the $z$-axis, respectively.

\section{Effective Hamiltonian and Multipartite Dynamics}

While sequence (\ref{sequence}) displays the basic concept of how
the cavity-laser mediated interaction is achieved between the atoms,
we still have to analyze this coupling quantitatively as to understand how
to control it in practice. For this purpose, we shall adiabatically
eliminate the intermediate states $\ket{a_1, \ldots, a_N, \bar{1}}$
and $\ket{e_1, \ldots, a_N,\bar{0}}, \ldots, \ket{a_1, \ldots, e_N,
$ $ \bar{0}}$ from the sequence (\ref{sequence}). This shall lead to
an effective Hamiltonian that describes the time evolution of $N$
atoms which interact with each other according to the simplified
sequence (\ref{sequence1}).

To outline this elimination process, let us first introduce the
short-hand notation
\begin{equation}\label{notation}
 \begin{array}{cc}
  & \ketv{1} \rightarrow \ketv{N+1} \rightarrow \ketv{0}
  \rightarrow \ketv{N+2} \rightarrow \ketv{2}
  \\
  & \hspace{3.8cm} \searrow
  \raisebox{-0.15cm}{$\ketv{N+3} \rightarrow \ketv{3}$}
  \\
  & \hspace{4.3cm} \vdots \hspace{1.3cm} \vdots
  \\
  & \hspace{4.3cm} \ketv{2N} \rightarrow \ketv{N},
  \end{array}
\end{equation}
for the composite states of $N$ identical atoms and the cavity,
which corresponds one-to-one to the states from sequence
(\ref{sequence}). With this notation, the W state (\ref{w-state1})
refers to the states $\ketv{1}, \ldots, \ketv{N}$, while the
cavity-mediated photon exchange is performed between the state
$\ketv{N+1}$ and (one of) the states $\ketv{N+2}, \ldots,
\ketv{2N}$, respectively.

For $N$ identical atoms, the evolution of the coupled
atoms-cavity-laser system is described by the Hamiltonian
\begin{eqnarray}\label{ham0}
H &=& \omega_C \, c^+ \, c + \sum_{i=1}^N \bigl(
      \omega_1 \ket{1}_i \bra{1} +
      \omega_E \ket{e}_i \bra{e} +
      \omega_A \ket{a}_i \bra{a} \nonumber
      \\
      & + & \left. \left[
         \Omega \, e^{-i \omega_L t} \ket{e}_i \bra{1} +
         g_i(t) \, c \, \ket{e}_i \bra{a} + h.c.
      \right] \right),
\end{eqnarray}
where the first term describes the cavity energy, with $c$ and $c^+$
being the annihilation and creation operators for a cavity photon
acting upon the Fock states $\ket{\bar{n}}$, and (the summation of)
the second term describes the chain of atoms and their interaction
with the laser and cavity. In this Hamiltonian, the interaction of
the $i-$th atom and the cavity is based on the time-dependent
coupling (\ref{coupling1}). In the summation of the second term,
moreover, each term contains the excitation energies $\omega_1$,
$\omega_E$, and $\omega_A$ which correspond to the atomic states
$\ket{1}$, $\ket{e}$, and $\ket{a}$, respectively.

In order to simplify the analysis of the time evolution associated
with the Hamiltonian (\ref{ham0}), we switch to the interaction
picture by using the unitary transformation \cite{jpb42, pra67}
\begin{small}
\begin{eqnarray}\label{picture}
U_I & = &
         \exp \left( -\im (\omega_1 + \omega_L) t \sum_{i=1}^N \ket{e}_i
               \bra{e} -\im \omega_1 t \sum_{i=1}^N \ket{1}_i \bra{1}
              \right) \nonumber
    \\
    & \times &
         \exp \left( -\im \omega_A \sum_{i=1}^N \ket{a}_i \bra{a}
                -\im \left[ \omega_L - \omega_A + \omega_1 \right]
                t \, c^+ \, c
               \right). \nonumber
\end{eqnarray}
\end{small}
In this picture, the Hamiltonian (\ref{ham0}) takes the form
\begin{eqnarray}\label{ham1}
H_I =  - \delta \, c^+ c + \sum_{i=1}^N H_i \quad \text{with}
       \hspace{1.5cm} && \\
H_i = \Delta \ket{e}_i \bra{e} +
      \left[
       \Omega \, \ket{e}_i \bra{1} +
       g_i(t) \, c \, \ket{e}_i \bra{a} + h.c.
      \right], && \nonumber
\end{eqnarray}
and where $\Delta = (\omega_E - \omega_1) - \omega_L$ and $\delta =
(\omega_L - \omega_C) - (\omega_A - \omega_1)$ refer to the two
off-resonance shifts (detunings) of the laser and cavity
frequencies, respectively, as depicted in Fig.~\ref{fig:2}(c). The
time evolution of the wavefunction then follows the Schr\"{o}dinger
equation
\begin{equation}\label{schrod_eq}
\im \frac{d \ket{\Psi}}{dt} = H_I \ket{\Psi},
\end{equation}
where we restrict the wavefunction $\ket{\Psi}$ to the ansatz
\begin{equation}\label{ansatz}
\ket{\Psi} = \exp \left(\im \frac{\Omega^2}{\Delta} t \right)
\sum_{i=0}^{2N} C_i(t) \ketv{i}, \quad C_i(0) = \delta_{i1}.
\end{equation}
For this ansatz, the Schr\"{o}dinger equation (\ref{schrod_eq}) gives rise
to a closed system of $2N+1$ equations ($i = 1, \ldots, N$)
\begin{small}
\begin{eqnarray}\label{solutions}
   \im \dot{C}_0(t) &=& \left( \frac{\Omega^2}{\Delta} -\delta \right) C_0(t)
             + \sum_{j=1}^N g_j(t) \, C_{N+j}(t), \nonumber
   \\
   \im \dot{C}_i(t) &=& \frac{\Omega^2}{\Delta} \, C_i(t) + \Omega \,
   C_{N+i}(t),
   \\[0.1cm]
   \im \dot{C}_{N+i}(t) &=& \left( \frac{\Omega^2}{\Delta}
          + \Delta \right) C_{N+i}(t)
          + \Omega \, C_{i}(t) + g_i(t) C_0(t) \, , \nonumber
\end{eqnarray}
\end{small}
and where the dot denotes the time derivative.

As explained above, the $N+1$ states $\ketv{0}$ and $\ketv{N+1},
\ldots, \ketv{2N}$ remain (almost) unpopulated if the atom-cavity
and atom-laser detuning satisfy the two conditions (\ref{condition})
and (\ref{condition1}), respectively. Therefore, in order to
separate these states from Eqs.~(\ref{solutions}), we utilize the
adiabatic elimination procedure which assumes an adiabatic behavior
of the functions $C_0(t)$ and $C_{N+1}(t), \ldots, C_{2N}(t)$ or, in
other words, that their time derivative vanishes to a good
approximation. Together with condition (\ref{condition1}) and
conditions
\begin{equation}\label{condition2}
   |\delta| \gg g(t), \quad
   |\delta \, \Delta| \gg \Omega^2, \quad
   |\delta \, \Delta| \gg g(t)^2 \, ,
\end{equation}
it is justified to eliminate $N+1$ equations from the system
(\ref{solutions}). Here, we shall omit further details of this
derivation for which the reader is refereed to the seminal paper
\cite{prl85}. The remaining $N$ (effective) equations for the
functions $C_i(t)$, take the form
\begin{equation}\label{solutions1}
\im \dot{C}_i(t) = \sum_{{j=1} \atop{(j \neq i)}}^N \frac{g_i(t) \,
                    g_j(t) \, \Omega^2}{\delta \, \Delta^2} C_j(t).
\end{equation}

Owning to the Eqs.~(\ref{solutions1}) for functions $C_i(t)$ and the
adiabatic behavior for functions $C_0(t)$ and $C_{N+1}(t)$,
$\ldots,$ $C_{2N}(t)$, the evolution of the overall state of the
atomic chain is given by the wavefunction
\begin{equation}\label{ansatz1}
\ket{\Phi} = \sum_{i=1}^{N} C_i(t) \ketv{i}
           = \exp \left( - \im \int_{-\infty}^{t} \hspace{-0.1cm}
                        H_{\text{eff}} \; ds \right) \ketv{1},
\end{equation}
which is associated with the effective Hamiltonian
\begin{equation}\label{ham2}
H_{\text{eff}} = \sum_{{i,j = 1} \atop{(i \neq j)}}^N
                  \frac{g_i(t) \; g_j(t) \: \Omega^2}{2 \, \delta \, \Delta^2}
                  \left( S_i^+ S_j^- + S_i^- S_j^+ \right),
\end{equation}
where $S_i^+ = \ket{1}_i \bra{a}$ and $S_i^- = \ket{a}_i \bra{1}$
denote the atomic two-photon excitation and de-excitation operators.
Obviously, this Hamiltonian (\ref{ham2}) is much simpler and
describes the effective atomic evolution (\ref{sequence1}) as
mediated by the combined laser and cavity fields in
Eqs.~(\ref{solutions1}). In order to summarize the steps before, we
have therefore found that the evolution of four-level atoms is
reduced to the evolution of effectively two-level atoms which
interact via a two-photon exchange in such a manner, that the
excited state $\ket{e}$ remains (almost) unpopulated.

When all the $N$ atoms have left the cavity, the wavefunction (\ref{ansatz1})
becomes in the limit $t \rightarrow +\infty$
\begin{equation}\label{wavefunction}
\ket{\Phi} = \exp \left( - \im \mathrm{M} \right) \ketv{1}, \quad
\mathrm{M} \equiv \int_{-\infty}^{+\infty} \hspace{-0.1cm}
H_{\text{eff}} \; ds \, ,
\end{equation}
where the matrix elements $\mathrm{M}_{ij} = \brav{i} \mathrm{M}
\ketv{j}$ are given by
\begin{equation}\label{matrix}
\mathrm{M}_{ii} = 0 \quad \text{and} \quad
\mathrm{M}_{ij} = \theta(\upsilon, |i-j| \, d) \quad \mbox{for} \quad i \neq j,
\end{equation}
with
\begin{equation}\label{theta}
   \theta(\upsilon, d) = \sqrt{\frac{\pi}{2}}
   \, \frac{\Omega^2 \, g^2_o \, w}{\delta \, \Delta^2 \, \upsilon} \,
   \exp \left( -\frac{{d \,}^2}{2 \, w^2} \right) \, .
\end{equation}
The latter expression (\ref{theta}) can be interpreted as the
asymptotic coupling for a pair of atoms that move with the same
velocity $\upsilon$ and are separated by the distance $d$ from each
other. The atomic evolution of the state (\ref{wavefunction}),
therefore, is completely characterized by the asymptotic coupling
(\ref{theta}) which depends on the two parameters $(\upsilon, d)$
once the frequency shifts and coupling constants $\delta$, $\Delta$,
$w$, $g_o$, and $\Omega$ are fixed by a particular experimental
setup.

\begin{figure*}
\begin{center}
\includegraphics[width=0.975\textwidth]{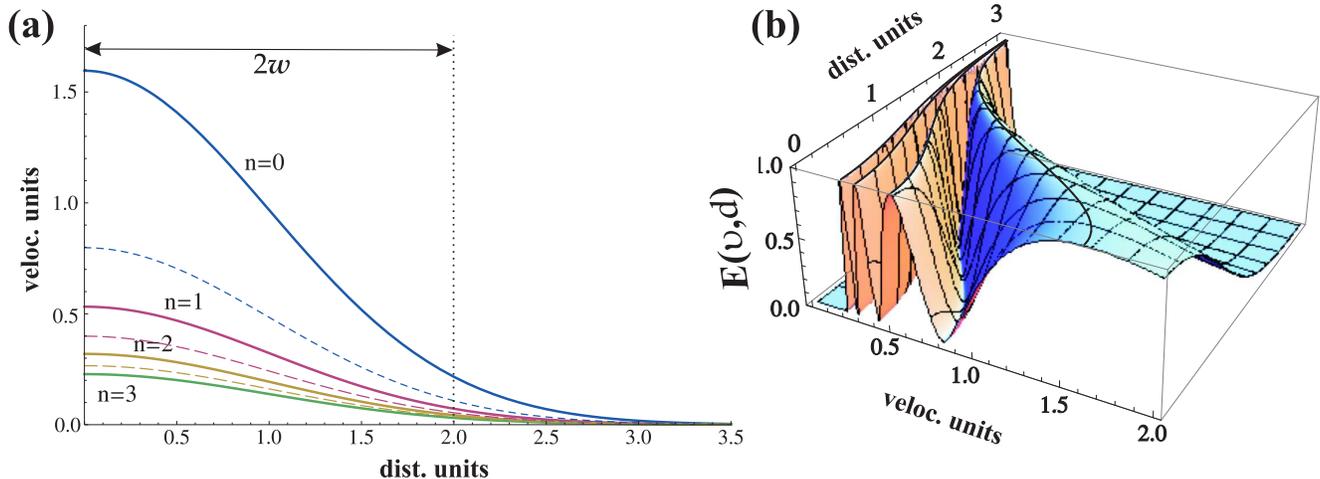} \\
\vspace{0.1cm}
\caption{(Color online) (a) Atomic velocities $\upsilon$ and
inter-atomic distances $d$ for which the initial product state of
two atoms $\ket{1_1, 0_2}$ becomes maximally entangled (solid
lines). The dashed lines, in contrast, indicate the $(\upsilon, d)$
pairs for which the atomic qubits remain disentangled. (b) Plot of
the von Neumann entropy $E(\upsilon, d)$ as a function of the atomic
velocity and distance. In all these figures, the velocities
$\upsilon$ are displayed in units of $\Omega^2 \, g^2_o \, w /
\delta \, \Delta^2$ and the distances $d$ in units of $w$.}
\label{fig:3}
\end{center}
\end{figure*}

Let us recall here that, after the atoms have left the cavity, their
population in the (electronic) state $\ket{a}$ is transferred
coherently back into the hyperfine state $\ket{0}$ by applying the
Raman pulses $L_2$. With this transfer, the wavefunction
(\ref{wavefunction}) then gives the entangled W-class state of $N$
hyperfine qubits
\begin{eqnarray}\label{wavefunction1}
&& \hspace{1.5cm} \ket{\Phi^\prime_N} =
                    \sum_{i=1}^N C_i(\upsilon, d) \ketvp{i}, \quad
                    \text{with}
\\
&& \ketvp{1} = \ket{1_1, \ldots, 0_N}, \; \ldots, \; \ketvp{N} =
               \ket{0_1, \ldots, 1_N}, \nonumber
\end{eqnarray}
and where the functions
\begin{equation}\label{functions}
C_i(+\infty) \equiv
C_i(\upsilon, d) = \bravp{i} \exp \left( - \im \mathrm{M} \right)
                   \ketvp{1}
\end{equation}
are obtained from the exponentiation of the hermitian operator
$\left( - \im \mathrm{M} \right)$. These functions can be computed
routinely for any number of atoms $N$, for instance, by
diagonalization of the matrix (\ref{matrix}).

The wavefunction (\ref{wavefunction1}), however, has not yet the
desired form of a W state (\ref{w-state}). In the next subsections,
we shall therefore discuss the properties of the W$_N$ states for
different values of $N$ and display those $\upsilon$ and $d$
parameters, for which the function $\ket{\Phi^\prime_N}$ is
equivalent (or close) to the desired W states.

\subsection{Two-partite entangled state}

For a chain of just two atoms $(N=2)$, the wavefunction (\ref{wavefunction1})
takes the simple form \cite{jpb42}
\begin{equation}\label{two-state}
\ket{\Phi^\prime_2} = \cos \theta(\upsilon, d) \,\ketvp{1} -
                      \im \sin \theta(\upsilon, d) \,\ketvp{2} \, .
\end{equation}
From this expression, we readily recognize that the two-partite
W$_2$ states
\begin{equation}\label{w2-state}
\ket{W_2^{\pm}} = \frac{1}{\sqrt{2}} \left( e^{\pm i \frac{\pi}{2}}
\ket{1_1, 0_2} + \ket{0_1, 1_2} \right)
\end{equation}
are obtained (up to a global phase factor) if the condition
$\theta(\upsilon, d) = (2n + 1) \, \pi/4$ is satisfied for some
integer $n$. For a fixed set of experimental parameters $\delta$,
$\Delta$, $w$, $g_o$, and $\Omega$, therefore, a maximum
entanglement is obtained only along the solid lines displayed in
Fig.~\ref{fig:3}(a) for $n = 0,1,2,3$. Obviously, the change between
maximally entangled (solid lines) and completely disentangled states
(dashed lines) happens more and more rapidly as the velocity of the
chain is decreased from a certain maximum value (namely, for $n=0$)
onwards.

Apart from understanding the dynamical parameters $(\upsilon, d)$
for which a maximum entanglement is achieved, it is important also
to know how sensitive these states are with regard to small
uncertainties in the velocity and inter-atomic distance. To analyze
this sensitivity, Fig.~\ref{fig:3}(b) displays the
\textit{von Neumann entropy} \cite{nc}
\begin{eqnarray}\label{entropy}
   E(\upsilon, d)
   & = &
   - \mbox{Tr} \left( \rho \, \log_2 \rho \right)
   \nonumber   \\
   & = &
   -\cos^2 \theta(\upsilon, d) \, \log_2 \left[ \cos^2 \theta(\upsilon, d)
   \right] \nonumber \\
   &   & \hspace*{1.0cm}
  -\sin^2 \theta(\upsilon, d) \, \log_2 \left[ \sin^2 \theta(\upsilon, d)
  \right] , \quad
\end{eqnarray}
for velocities and distances satisfying $\theta (\upsilon, d) < 2 \,
\pi$, and where $\rho = \mbox{Tr}_2 \left( \ket{\Phi^\prime_2}
\bra{\Phi^\prime_2} \right)$ denotes the reduced density operator
with regard to the second hyperfine qubit. As expected, the maximal
values of the von Neumann entropy, i.e., $E(\upsilon, d) = 1$, are
obtained along the lines which are displayed in Fig.~\ref{fig:3}(a).
Moreover, the least rapid variation in the maxima occurs along the
$n=0$ line and for rather small inter-atomic distances. For small
velocities or some larger distance of the atoms, in contrast, a good
control of the entanglements of the $\ket{\Phi^\prime_2}$ states
becomes more and more difficult.

Fig.~\ref{fig:3}(a) shows that an entanglement between the atoms
occurs even for inter-atomic distances which are larger than $2w$,
i.e.\ twice the cavity waist. In a high finesse cavity, indeed, the
Gaussian profile (\ref{coupling}) approximates quite well the
intra-cavity field and, thus, it is possible to generate an
entangled state even for the atomic separation $d > 2w$. In
practice, however, the cavity relaxation and the spontaneous decay
of the atoms introduce certain limitations on the distance between
the atoms, beyond which it is not possible to generate the entangled
state (\ref{w2-state}). In order to estimate this limitation, we
consider the condition \cite{hc}
\begin{equation}
N \, g_o^2 \, \exp \left[ - 2 \, z^2 / w^2 \right] / \left( \kappa
\, \gamma \right) > 1
\end{equation}
which ensures that $N$ atoms couple strongly to the cavity field
and, therefore, implies the validity of the effective evolution
(\ref{ham2}). Here, $\kappa$ and $\gamma$ denote the cavity loss
rate and the atomic decay rate, respectively. For $N=2$, the above
condition bounds the atomic coordinate to the interval $z_- < z <
z_+$ with
\begin{equation}
z_\pm \;=\; \pm \, w \,
\sqrt{\frac{\ln[ 2 \, g_o^2 / \left( \kappa \, \gamma \right) ]}{2}}.
\end{equation}
Owning to these boundaries, therefore, the distance $d$ between two
atoms must satisfy
\begin{equation}
\frac{d}{w} < \sqrt{2 \, \ln[ 2 \, g_o^2 / \left( \kappa \, \gamma
\right) ]} = \frac{z_+}{w} - \frac{z_-}{w} \, .
\end{equation}
For the typical atom-cavity parameters \cite{njp10}: $\{g_o, \kappa,
\gamma \} \,=\, 2 \pi \times \{10, 0.4, 2.6 \}$ MHz, this codition
implies the limitation $d < 3.243 \, w$. We note that this
estimation agrees well with the solid lines from Fig.~\ref{fig:3}(a)
since, for $d > 3.2 \, w$, the atomic velocity becomes so small that
it would prevent any experimental implementation of our scheme.

\begin{figure}
\begin{center}
\includegraphics[width=0.45\textwidth]{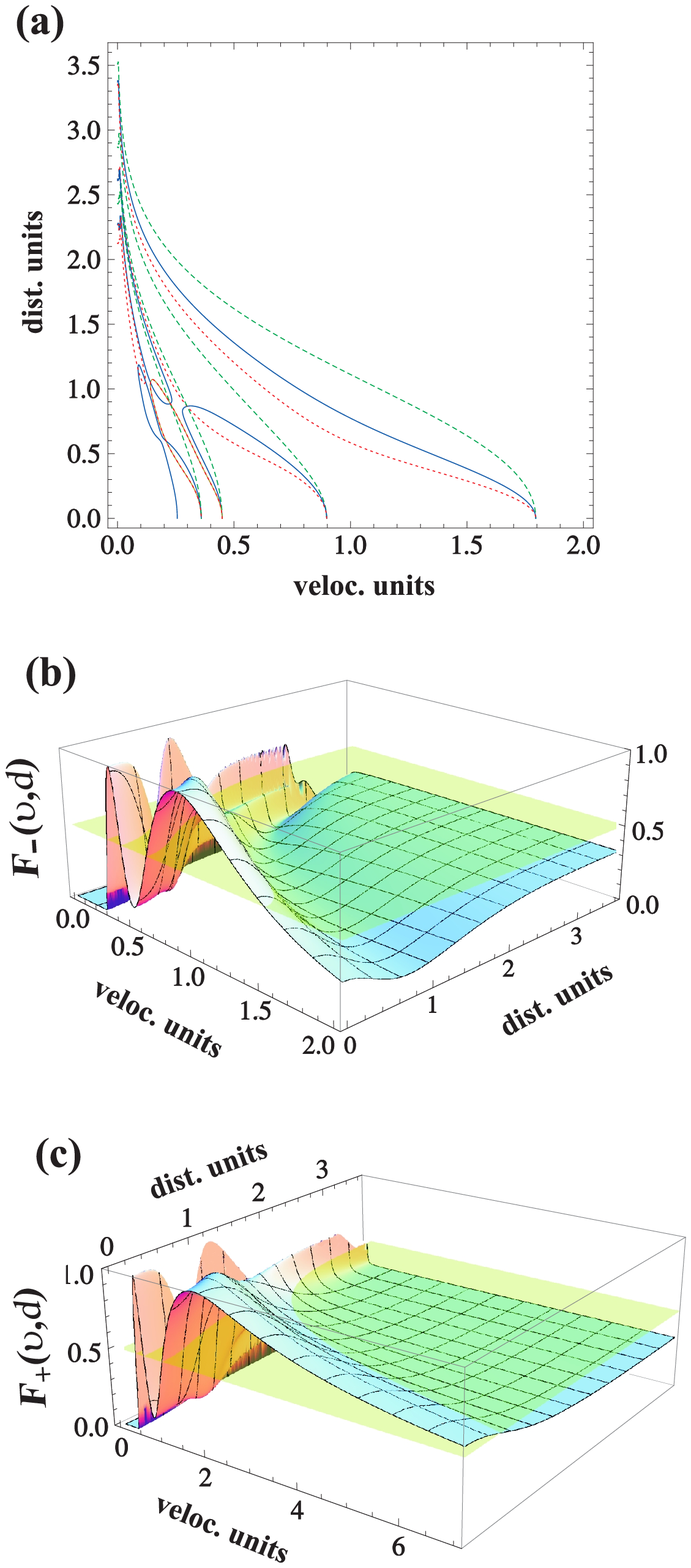} \\
\vspace{0.1cm}
\caption{(Color online) (a) Lines along which the moduli $|C_1
(\upsilon, d)|$ (solid), $|C_2 (\upsilon, d)|$ (dashed), and $|C_3
(\upsilon, d)|$ (dotted) are equal to $1 / \sqrt{3}$. These lines
correspond to velocities (\ref{condition3}) with $n = 0,1$ and $m =
1,2$ (in the limit $d \rightarrow 0$). (b) Two maxima in the
fidelity $F_{-}(\upsilon, d)$ obtained for velocities
(\ref{condition3}) with $n = 0,1$ and $m = 1,2$ (in the limit $d
\rightarrow 0$). For guidance of the eyes, the semi-transparent
layer displays a constant value $F_{-}(\upsilon, d) = 0.5$. (c) The
same as in figure (b) but for the fidelity $F_{+}(\upsilon, d)$.
Again, all velocities $\upsilon$ are displayed in units of $\Omega^2
\, g^2_o \, w / \delta \, \Delta^2$ and the distances $d$ in units
of $w$.}
\label{fig:4}
\end{center}
\end{figure}

\subsection{Tree-partite W state}

For a chain of three atoms $(N=3)$, the wavefunction (\ref{wavefunction1})
takes the form
\begin{equation}\label{three-state}
\ket{\Phi^\prime_3} = \sum_{i=1}^3 C_i(\upsilon, d) \ketvp{i},
\end{equation}
with
\begin{eqnarray}\label{functions1}
C_1(\upsilon, d) & = & \frac{- \xi^3 \, \lambda_{-} +
                       \sqrt{8 + \xi^6}
               \left( \lambda_{+} + 2 \, e^{i \, \kappa} \right)}
                       {4 \, \sqrt{8 + \xi^6}} \, e^{-i \, \zeta}, \nonumber
\\
C_2(\upsilon, d) & = & - \frac{\lambda_{-}}{\sqrt{8 + \xi^6}}
                         \, e^{-i \, \zeta},
\\
C_3(\upsilon, d) & = &  \frac{- \xi^3 \, \lambda_{-} +
                        \sqrt{8 + \xi^6}
            \left( \lambda_{+} - 2 \, e^{i \, \kappa} \right)}
                        {4 \, \sqrt{8 + \xi^6}} \, e^{-i \, \zeta} \, ,
\nonumber
\end{eqnarray}
and where we used the notation $\xi = \exp \left[ - {d \,}^2 / (2 \,
w^2) \right]$ and
\begin{eqnarray}
\lambda_{\pm} & = & \exp \left[ i \, 2 \, \xi \, \chi \sqrt{8 +
                    \xi^6} \right] \pm 1 \, ,
\nonumber \\
\chi          & = & \sqrt{\frac{\pi}{8}} \, \frac{\Omega^2 \, g^2_o
\, w}
                    {\delta \, \Delta^2 \, \upsilon} \, ,
\nonumber \\
\kappa        & = & \xi \, \chi \left( 3 \, \xi^3 + \sqrt{8 + \xi^6} \right),
\nonumber \\
\zeta         & = & \xi \, \chi \left( \xi^3 + \sqrt{8 + \xi^6} \right) \, .
\nonumber
\end{eqnarray}
In order to obtain the state W$_3$ from wavefunction
$\ket{\Phi^\prime_3}$, we have to determine those pairs $(\upsilon,
d)$ for which the equations
\begin{equation}\label{ansatz2}
|C_1 (\upsilon, d)| = |C_2 (\upsilon, d)| = |C_3 (\upsilon, d)| =
\frac{1}{\sqrt{3}}
\end{equation}
are fulfilled. In Fig.~\ref{fig:4}(a), we displayed the
corresponding lines for which the moduli $|C_1(\upsilon, d)|$
(solid), $|C_2(\upsilon, d)|$ (dashed), and $|C_3(\upsilon, d)|$
(dotted) are equal to $1 / \sqrt{3}$. The requested W$_3$ states are
obtained for those $(\upsilon, d)$ pairs, for which all three types
of lines intersect with each other.

As seen from Fig.~\ref{fig:4}(a), however, the lines for the (moduli
of the) functions $C_i(\upsilon, d)$ intersect only if the
inter-atomic distance vanishes. In order to determine the
corresponding velocities, we first observe that for $d \rightarrow
0$ ($\xi \rightarrow 1$), the wavefunction (\ref{three-state})
becomes
\begin{equation}
e^{-i \, 4 \chi} \frac{1 + 2 \, e^{i \, 6 \chi }}{3} \ketvp{1} \, +
\, e^{-i \, 4 \chi} \frac{1 - e^{i \, 6 \chi }}{3} \left( \ketvp{2}
+ \ketvp{3} \right) \, . \nonumber
\end{equation}
This expression can be readily cast into the $W_3$ form
\begin{equation}\label{w3-state}
\ket{W_3^{\pm}} = \frac{1}{\sqrt{3}} \left( e^{\pm i \frac{2
\pi}{3}} \ketvp{1} + \ketvp{2} + \ketvp{3} \right)
\end{equation}
if $\chi = (3 \, n + m) \, \pi /9$ or, equivalently, if the velocity
takes the values
\begin{equation}\label{condition3}
\upsilon = \sqrt{\frac{\pi}{8}} \, \frac{\Omega^2 \, g^2_o \, w}
{\delta \, \Delta^2} \frac{9}{\pi (3 \, n + m)} \, ,
\end{equation}
with $m=1,2$ and $n$ being an integer. To summarize, the vanishing
inter-atomic distance along with velocities (\ref{condition3}) are
both necessary to obtain the W$_3$ states due to wavefunction
(\ref{three-state}). According to our scheme, however, the atoms are
separated by a macroscopic distance which is non-negligible with
regard to the cavity waist. Therefore, we may determine the
parameter region $(\upsilon, d)$ with the non-zero inter-atomic
distance, for which the two fidelities \cite{nc}
\begin{eqnarray}\label{fidelity}
F_{\pm}(\upsilon, d)
                    & = & | \langle W_3^\pm | \Phi^\prime_3 \rangle |^2
                    \\
                    & = & \frac{1}{3} \left| e^{\mp i \, \frac{2 \pi}{3}} \,
                    C_1 (\upsilon, d) + C_2 (\upsilon, d) + C_3 (\upsilon, d)
                                      \right|^2, \nonumber
\end{eqnarray}
between the states $\ket{W_3^\pm}$ and $\ket{\Phi^\prime_3}$ are
larger than the threshold value of $1/2$. These fidelities
$F_{-}(\upsilon, d)$ and $F_{+}(\upsilon, d)$ are displayed in
Figs.~(\ref{fig:4})(b) and (c), together with a semitransparent
plane in order to delimit the regions for which $F_{\pm}(\upsilon,
d) \geq 0.5$. While the maximum values $F_{\pm}(\upsilon, d) = 1$
are obtained only for a few velocities and vanishing inter-atomic
distance, there are still $(\upsilon, d)$ regions (with non-zero
distance) for which the fidelities become reasonably close to the
maximal value. Note, moreover, that the region with $F_{+}(\upsilon,
d) \geq 0.5$ is notably larger than those with $F_{-}(\upsilon, d)
\geq 0.5$. We conclude, therefore, that from the experimental
perspective it might be preferable to generate the $\ket{W_3^+}$
state between three hyperfine qubits by means of the suggested
scheme.

\begin{figure}
\begin{center}
\includegraphics[width=0.45\textwidth]{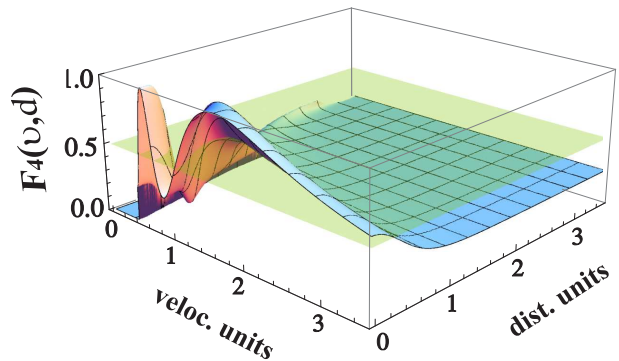} \\
\vspace{0.1cm}
\caption{(Color online) Fidelity $F_4(\upsilon, d)$ for the
generation of the W$_4$ state as a function of the velocity and the
inter-atomic distance. Again, the maximum value $F_4(\upsilon, d) =
1$ is obtained only for vanishing distance ($d = 0$) and velocities
(\ref{condition4}) with $n = 0,1$. The semi-transparent plane with
$F_4(\upsilon, d) = 0.5$ is plotted to guide the eyes; the units are
the same as in Figs.~\ref{fig:3} and \ref{fig:4}.}
\label{fig:5}
\end{center}
\end{figure}

\subsection{Four-partite W state}

For a chain of four atoms $(N=4)$, the wavefunction (\ref{wavefunction1})
can be written as
\begin{equation}\label{four-state}
\ket{\Phi^\prime_4} = \sum_{i=1}^4 C_i(\upsilon, d) \ketvp{i} \, .
\end{equation}
In contrast to $N=2$ or $N=3$, however, the expressions for
$C_i(\upsilon, d)$ become rather bulky now and are not displayed
here. Recall from the previous subsection that the wavefunction
$\ket{\Phi^\prime_3}$ produced the $\ket{W_3^{\pm}}$ states only for
vanishing distances and velocities (\ref{condition3}). In this
subsection, therefore, we proceed in a similar fashion and consider
the wavefunction (\ref{four-state}) with the vanishing inter-atomic
distance $(d \rightarrow 0)$
\begin{equation}
e^{-i \, 4 \chi} \frac{1 + 3 \, e^{i \, 8 \chi}}{4} \ketvp{1} +
e^{-i \, 4 \chi} \frac{1 - e^{i \, 8 \chi}}{4} \sum_{i=2}^4
\ketvp{i}. \nonumber
\end{equation}
From this expression, the state
\begin{equation}\label{w4-state}
\ket{W_4} = \frac{1}{2} \left( e^{i \, \pi} \ketvp{1} + \ketvp{2} +
\ketvp{3} + \ketvp{4} \right)
\end{equation}
is readily produced if $\chi = \pi (2 \, n + 1) /8$ or,
equivalently, if the velocity takes the values
\begin{equation}\label{condition4}
\upsilon = \sqrt{\frac{\pi}{8}} \,
           \frac{\Omega^2 \, g^2_o \, w}{\delta \, \Delta^2} \,
       \frac{8}{\pi \, (2 \, n + 1)}.
\end{equation}

More generally, Fig.~(\ref{fig:5}) displays the fidelity
$F_4(\upsilon, d) = | \langle W_4 | \Phi^\prime_4 \rangle |^2$ for
the generation of the W$_4$ states as a function of the velocity and
the inter-atomic distance due to the wavefunction
(\ref{four-state}). Analogue to the last subsection, the maximum
fidelity $F_4(\upsilon, d) = 1$ is obtained only for zero distance
($d = 0$) and velocities that fulfill the condition
(\ref{condition4}). For non-zero distances, nevertheless, there is
one broad parameters region for which the W$_4$ state
(\ref{w4-state}) can be generated with a reasonable hight fidelity.

\subsection{$N \geq 5$ partite W state}

For any other number $N \ge 5$ of atoms in the chain, the functions
$C_1 (\upsilon, d), \ldots, C_N (\upsilon, d)$ can still be computed
from the formula (\ref{functions}), and the wavefunction
$\ket{\Phi^\prime_N}$ can be further analyzed with regard to
$(\upsilon, d)$ region, for which the corresponding $W_N$ state is
produced most reliably. In order to generate such state, according
to the definition (\ref{w-state}), the conditions
\begin{equation}\label{condition5}
|C_1 (\upsilon, d)| = \ldots = |C_N (\upsilon, d)| =
\frac{1}{\sqrt{N}}
\end{equation}
need to be fulfilled. By performing numerical analysis, however, it
turns out that the Eqs.~(\ref{condition5}) cannot be fulfilled for
any choice of the velocity and inter-atomic distance and, therefore,
the fidelity $| \langle W_N | \Phi^\prime_N \rangle |^2$ is always
smaller than unity.

Nevertheless, we can display this fidelity as a function of
$\upsilon$ and $d$ and determine the region where it takes the
highest value. In order to proceed so, however, we still need to
specify the \textit{reference} state $\ket{W_N}$, which we are
looking in the form
\begin{equation}\label{wn-state0}
\ket{W_N} = \frac{1}{\sqrt{N}} \left( e^{i \, \phi} \ketvp{1} +
\sum_{i=2}^N \ketvp{i} \right)
\end{equation}
with an unknown phase $\phi$. The form of this state has been chosen
in line with the previously obtained W states (\ref{w2-state}),
(\ref{w3-state}), and (\ref{w4-state}). In order to calculate the
unknown phase $\phi$ in (\ref{wn-state0}), let us first consider the
W-class state
\begin{eqnarray}\label{n-state}
\ket{\widetilde{\Phi}^\prime_N}
   & = & e^{i \, 4 \chi} \, \sum_{k=1}^N  C_k(\upsilon, 0) \ketvp{k} \\
   & = & \sum_{k=1}^N \frac{1 + (\delta_{k 1}N - 1)
                      \exp \left( \im \, 2 N \chi \right)}{N}
                      \ketvp{k}, \nonumber
\end{eqnarray}
where the velocity $\upsilon$ is such that it makes the expressions
\begin{equation}\label{expressions}
\left| |C_1(\upsilon, 0)| - \frac{1}{\sqrt{N}} \right|, \; \ldots,
\; \left| |C_N(\upsilon, 0)| - \frac{1}{\sqrt{N}} \right|
\end{equation}
minimal. It can be straightforwardly shown that all the expressions
(\ref{expressions}) are minimized for the values $\chi = \pi (2 \, n
+ 1)/ (2 \, N)$ or, equivalently, for velocities
\begin{equation}\label{condition6}
\upsilon =  \sqrt{\frac{\pi}{8}} \,
           \frac{\Omega^2 \, g^2_o \, w}{\delta \, \Delta^2} \,
       \frac{2N}{(2 \, n + 1)} \, .
\end{equation}
Substituting this value for $\chi$, the state (\ref{n-state}) then becomes
\begin{equation}\label{n-state1}
\ket{\widetilde{\Phi}^\prime_N} = \frac{2 - N}{N} \ketvp{1} +
\frac{2}{N} \sum_{i=2}^N \ketvp{i}.
\end{equation}
For a vanishing inter-atomic distance, therefore, the state
$\ket{\widetilde{\Phi}^\prime_N}$ with velocities (\ref{condition6})
gives the best approximation to the W$_N$ state.

\begin{figure}
\begin{center}
\includegraphics[width=0.45\textwidth]{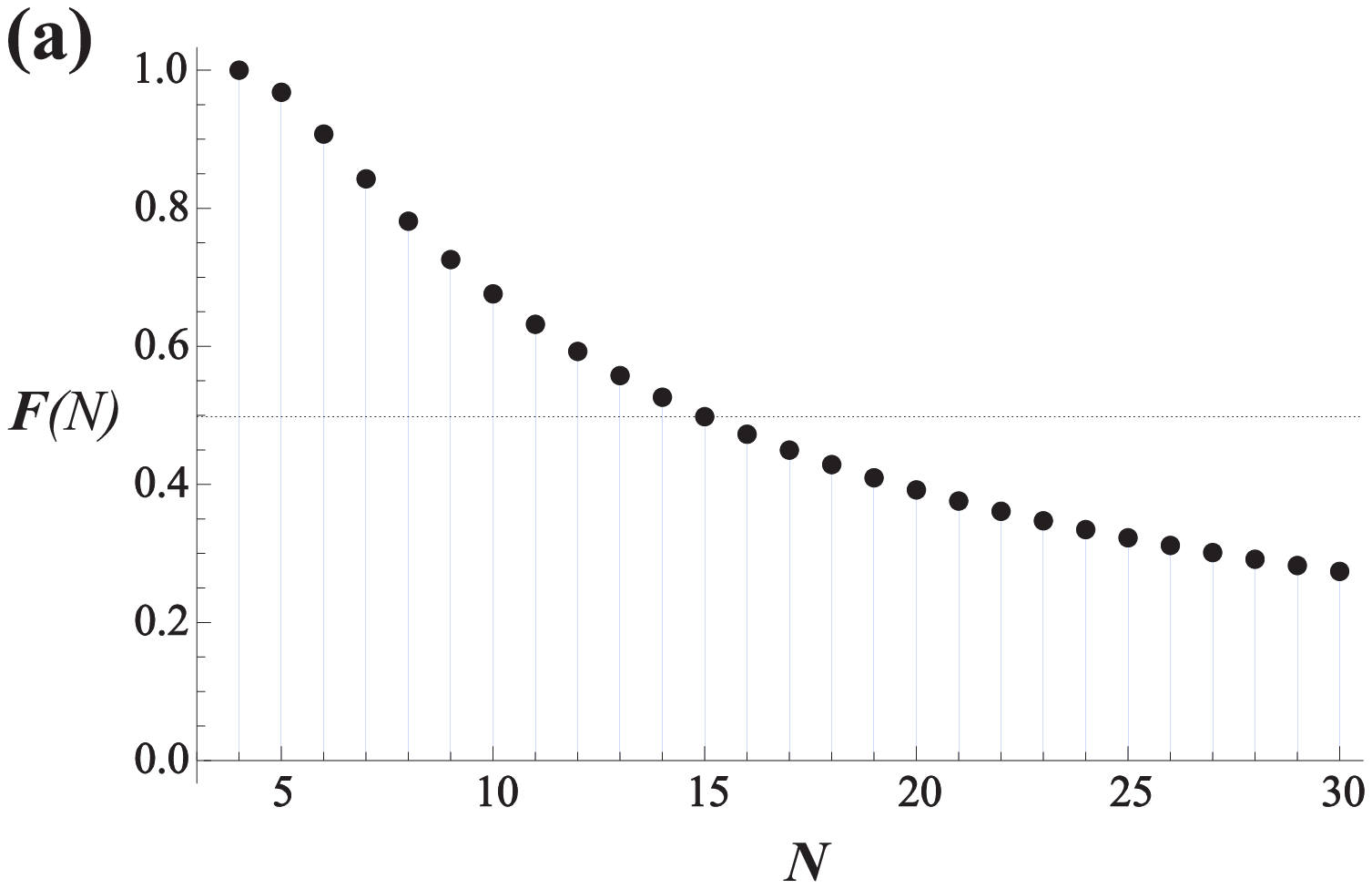} \\
\vspace{0.5cm}
\includegraphics[width=0.45\textwidth]{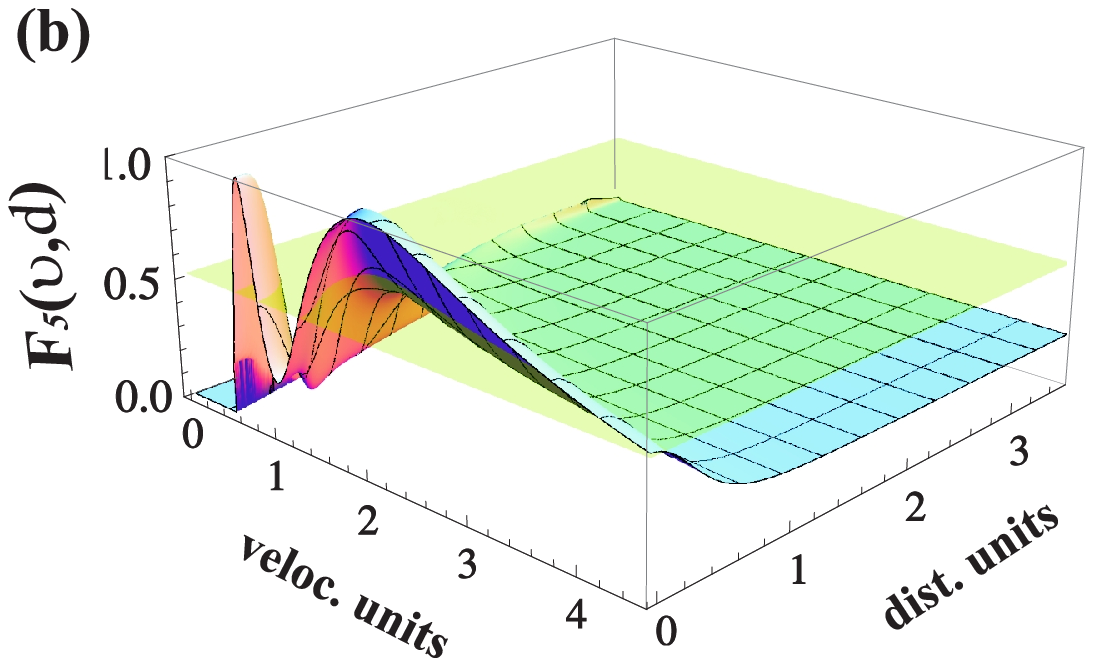} \\
\vspace{0.1cm}
\caption{(Color online) (a) The fidelity $F(N) = | \langle W_N
|\widetilde{\Phi}^\prime_N \rangle |^2$ has its maximum value $F(N)
= 1$ for $N=4$ and decreases monotonically as the number of atoms
increases. For $N > 15$, it falls below the threshold $F(N) = 1/2$
(dotted line). (b) Fidelity $F_5(\upsilon, d)$ for the production of
the $\ket{W_5}$ state due to $\ket{\Phi^\prime_5}$ as a function of
the velocity and the interatomic distance. Similar as in
Fig.~\ref{fig:5}, it reaches its maxima for $d = 0$ and velocities
(\ref{condition6}) with $n = 0,1$. The units are the same as in
Figs.~\ref{fig:3} and \ref{fig:4} above.}
\label{fig:6}
\end{center}
\end{figure}

As we explained in the beginning of this subsection, there are no
such $(\upsilon, d)$ pairs for which the Eqs.~(\ref{condition5}) can
be fulfilled. However, we found the state (\ref{n-state1}) which
gives the best approximation to the W$_N$ state (\ref{wn-state0}).
By comparing the states $\ket{W_N}$ with
$\ket{\widetilde{\Phi}^\prime_N}$ for $N=4$, we find that the phase
$\phi$ is equal to $\pi$ and, therefore, the reference state becomes
\begin{equation}\label{wn-state}
\ket{W_N} = \frac{1}{\sqrt{N}} \left( e^{i \, \pi} \ketvp{1} +
\sum_{i=2}^N \ketvp{i} \right), \quad N \geq 4.
\end{equation}
In order to understand how well the state (\ref{n-state1})
approximates the above state (\ref{wn-state}), in
Fig.~\ref{fig:6}(a) we displayed the fidelity $F(N) = | \langle W_N
| \widetilde{\Phi}^\prime_N \rangle |^2$. As seen from this figure,
the fidelity has its maximum value $F(N) = 1$ for $N=4$ and
decreases monotonically as the number of atoms is increases in the
chain. The fidelity drops below the threshold $F(N) = 1/2$ for $N >
15$. We therefore conclude that the state (\ref{n-state1})
approximates reasonably well the reference state for at least $4 < N
< 15$.

Having specified the reference state (\ref{wn-state}), we can
evaluate the fidelities
\begin{equation}\label{fidelity1}
F_N(\upsilon, d) = | \langle W_N | \Phi^\prime_N \rangle |^2; \quad
5 \leq N < 15
\end{equation}
as functions of the velocity and the inter-atomic distance. In
Fig.~\ref{fig:6}(b), for instance, we display the fidelity
(\ref{fidelity1}) for $N=5$. According to this figure, moreover, the
fidelity reaches its maxima $F_5(\upsilon, d) = F(5) \approx 0.97$
for $d = 0$ and velocities that satisfy the condition
(\ref{condition6}) with $n = 0,1$. Let us note here that the typical
spacing between two neighbored potentials wells (sites) of an
optical lattice is in the sub-micrometer range \cite{prl95, prl98,
njp10}. As seen from Fig.~\ref{fig:6}(b), this typical spacing is
comparable to the inter-atomic distance for which the fidelity $F_5
(\upsilon, d) \approx 0.9$ is reasonable high and where the typical
cavity waist ($w = 20$ $\mu$m) has been considered as the distance
units. The recent developments in cavity QED, therefore, make it
possible to generate the W$_5$ state by means of the proposed
scheme. If we compare, however, the $(\upsilon, d)$ regions for
which the fidelities $F_\pm (\upsilon, d)$
[Fig.~\ref{fig:4}(b)-(c)], $F_4 (\upsilon, d)$ [Fig.~\ref{fig:5}],
and $F_5 (\upsilon, d)$ [Fig.~\ref{fig:6}(b)] are higher than the
threshold value of $1/2$, we also conclude that these regions become
smaller as the number of atoms (in the chain) increases.

\begin{figure*}
\begin{center}
\includegraphics[width=0.975\textwidth]{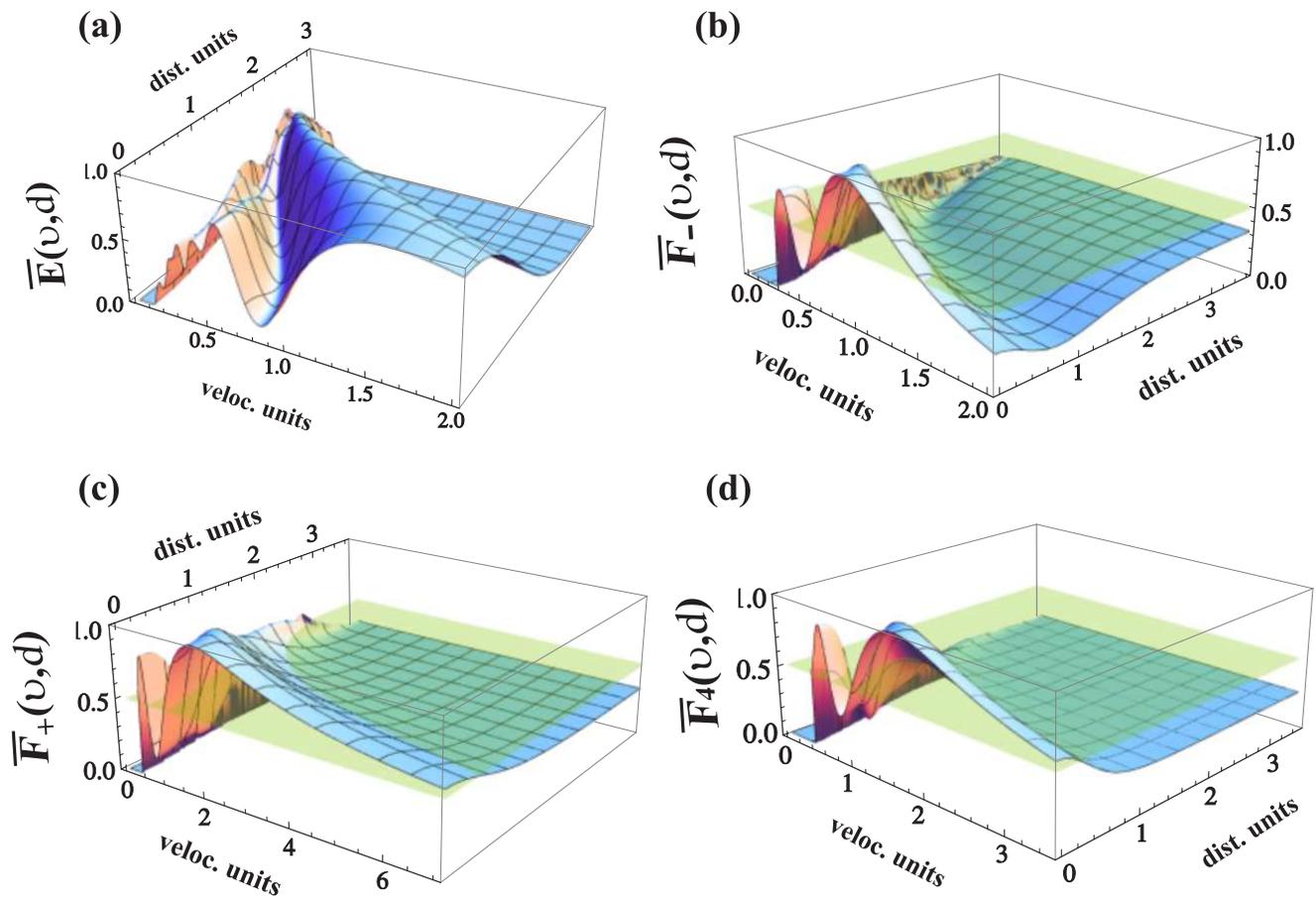} \\
\vspace{0.1cm}
\caption{(Color online) Entanglement and fidelity measures averaged
over $20$ randomly chosen uncertainties $\var{d}$ and
$\var{\upsilon}$ of the inter-atomic distance and the velocity,
respectively. (a) Von Neuman entropy $\overline{E}(\upsilon, d)$,
(b) fidelities $\overline{F}_{-}(\upsilon, d)$ and (c)
$\overline{F}_{+}(\upsilon, d)$ for a chain of three atoms, and (d)
$\overline{F}_4(\upsilon, d)$ for a chain of four atoms. The
uncertainties for the distance and velocities are chosen from the
intervals $[0, \, 0.2 \, \overline{d}]$ and $[0, \, 0.1 \,
\overline{\upsilon}]$, respectively (see text for further details).}
\label{fig:7}
\end{center}
\end{figure*}

\section{Remarks on the Implementation of the Proposed Scheme}

In our discussions so far, we have always assumed that the velocity
and the distance of the atoms in the chain, i.e.~their position
within the optical lattice, can be controlled exactly. With this
assumption in mind, the atom-cavity coupling was described by
expression (\ref{coupling1}). This assumption, however, neglects the
transversal components of the cavity field as well as the
oscillations of the atoms within the potential well due to their
finite temperatures, which include both the axial (along the
$z$-axis) and radial (along the $x , y$-axes) oscillations [see
Fig.~\ref{fig:2}(a)]. This additional motion gives rise to a
dispersion of the atomic positions and velocities and, thus, leads
to uncertainties in selecting the dynamical parameters in our model.

Obviously, any significant uncertainty in the parameters $\{
\upsilon, d, \delta, \Delta, g_o, \Omega \}$ will influence the
generation of the desired W entangled states. According to our
scheme, however, these entangled states are produced when all the
atoms have left the cavity. Instead of understanding these
parameters as \textit{exact}, therefore, they should refer to the
mean values and we need to analyze how small (but realistic)
variations in these parameters affect the final state of the atomic
chain. For instance, the radial oscillations of the atoms lead to
the mean value of the vacuum Rabi frequency $\overline{g}_o$ and
axial oscillations to the mean values of the inter-atomic distance
$\overline{d}$ and velocity $\overline{v}$, respectively. Axial
oscillations affects also the initial position $z^o_{i}$ and
velocity $\upsilon_{i}$ of each atom inside the lattice and result
in uncertainties $\var{d_i} = \overline{d} - |z^o_{i+1} - z^o_{i}|$
and $\var{\upsilon_i} = \overline{\upsilon} - \upsilon_i$, where $i
= 1, \ldots, N$. Therefore, the plots $E(\upsilon, d)$,
$F_\pm(\upsilon, d)$, $F_4(\upsilon, d)$, and $F_5(\upsilon, d)$
from Figs.~(\ref{fig:3})-(\ref{fig:6}) should be re-calculated as
function of the mean values $\overline{\upsilon}$ and $\overline{d}$
and their corresponding uncertainties, respectively.

In order to determine realistic uncertainties for the distance and
velocity of the atoms in the chain, we first mention that recent
cavity QED experiments allow to position the atoms relative to the
cavity antinode with a precision of $\sim 0.1$ $\mu$m by utilizing
an additional dipole trap acting along the cavity $y-$axis
\cite{prl95}. When compared to the typical cavity wavelength ($\sim
0.8$ $\mu$m), such positioning precision leads to the spatial
dispersion which, in turn, yields the mean value $\overline{g}_o
\approx 0.75 \, g_o$ that is still good enough for our purposes [see
Eq.~(\ref{condition2})]. Moreover, the same spatial dispersion
implies upper bounds for the uncertainties $| \hspace{-0.15cm}
\var{d} / \overline{d}| \lesssim 0.2$ and $|\hspace{-0.15cm}
\var{\upsilon} / \overline{\upsilon}| \lesssim 0.1$, if compared
with the typical spacing ($\sim 0.5$ $\mu$m) between two neighbored
potential wells of an optical lattice and the typical atomic
velocities ($\sim 0.5$ m/s) along the lattice axis.

For a further analysis of how reliably a given (experimental) setup
will generate a particular W state, in Fig.~\ref{fig:7} we display
the (mean) functions $\overline{E}(\upsilon, d)$,
$\overline{F}_\pm(\upsilon,d)$, and $\overline{F}_4(\upsilon, d)$ by
calculating their average for a certain spread of parameters. For
each subfigure ~\ref{fig:7}(a)-(d), we have randomly chosen $20$
uncertainties $\var{d}$ and $\var{\upsilon}$ from the intervals $[0,
\, 0.2 \, \overline{d}]$ and $[0, \, 0.1 \, \overline{\upsilon}]$,
respectively. By comparing the Figs.~\ref{fig:3}(c) and
\ref{fig:7}(a) it can be seen that the von Neumann entropy, for
instance, is reduced considerably for its sharp maxima ($n = 3$) and
that it remains almost the same around the broad maxima ($n = 0$).
Similarly, the mean fidelities which are displayed in
Figs.~\ref{fig:7}(b)-(d), are considerably reduced for their sharp
maxima. These $(\upsilon, d)$ regions for the velocity and
inter-atomic distance in the atomic chain are, therefore, less
useful for any practical implementation and only the $(\upsilon, d)$
regions which correspond to the broad maxima of the von Neumann
entropy and fidelities, are relevant for the generation of entangled
W states by means of the proposed scheme.

Finally, we assumed in our treatment that the (center-of-mass)
position of each atom is not a quantum variable but described
classically by the vector $\vec{r}_i(t) = \{ 0,0, z_i^o + \upsilon
\, t \}$. Obviously, such an assumption excludes several important
effects on the atomic motion that arise due to quantization of the
cavity field. For example, the correlations between the internal
dynamics of the atoms and their transverse (center-of-mass) position
may lead to an additional source of decoherence and disentanglement
in the effective evolution (\ref{ham2}), if the atom-cavity system
is embedded in a realistic reservoir \cite{epjd54}. In our scheme
however, the external potential that is created by the optical
lattice dominates the kinetic energy associated with the atomic
momentum. Therefore, any correlations which are induced by the
mechanical effects of cavity on atoms are strongly suppressed and
have been neglected in our treatment.

\section{Summary and Outlook}

A scheme is proposed to generate the entangled W states for a chain
of $N$ four-level atoms that are equally separated and conveyed
through an optical cavity by means of an optical lattice. This
scheme is based on the cavity-laser mediated interaction between the
atoms which are separated by a macroscopic distance, and works in a
completely deterministic way for qubits encoded by two hyperfine
levels of the atoms. Only two parameters, namely the velocity
$\upsilon$ of the chain and the inter-atomic distance $d$, determine
the effective interaction among the atoms and, thus, the degree of
entanglement that is obtained for the overall chain. The asymptotic
coupling (\ref{theta}) that completely characterizes the atomic
evolution, tells explicitly how the degree of entanglement depends
on these two parameters. The purpose of this work is to understand
the state evolution of the atomic chain and how it can be utilized
to generate the entangled W states. For chains consisting of $N =
2,3,4$ and $5$ atoms, Figs.~(\ref{fig:3})-(\ref{fig:6}) display the
von Neumann entropy and the fidelities as functions of the velocity
and inter-atomic distance. For $5 \leq N < 15$, moreover, we
suggested the reference state (\ref{wn-state}) which is approximated
by the wavefunction $\ket{\Phi^\prime_N}$ with a high fidelity. In
view of the recent developments in cavity QED, moreover, we have
also analyzed and discussed the proposed scheme with regard to
sensitivity in the formation of desired entanglement due to
uncertainties in the atomic motion.

For two or more atoms, the generation of entanglement by means of a
(detuned) optical cavity has been investigated in several papers
\cite{pra67, prl85, pra65}. All these studies, however, relied on
the \textit{small sample approximation} in which the separation of
the atoms is considered to be negligible when compared with the
cavity waist. Only recently \cite{pra71, pra68, pra75a}, the
atom-cavity coupling (\ref{coupling}) has been exploited in more
detail in order to suggest various entanglement schemes within
cavity QED. In the work by Amniat-Talab et al., for instance, a
scheme was proposed in which two atoms were coupled sequentially to
a resonant cavity and where a position-dependent coupling is used to
drive a STIRAP-type process in order to reduce the losses due to
atomic and cavity decays. Moreover, the scheme by Marr et al.\ is
also based on a STIRAP-type process and describes an adiabatic
evolution of a product state of two atoms which are coupled
simultaneously to a detuned cavity. The success of this scheme,
however, relies strongly on the ability to detect the photons which
leak through the cavity mirrors with an efficiency close to one. In
both schemes, therefore, the atomic velocities and inter-atomic
separation are used to control the accuracy of a STIRAP-type
process, in contrast to our approach, in which these parameters are
utilized to control the degree of entanglement.

Our proposed scheme might be suitable also for ion-cavity
experiments in which $N$ trapped ions interact simultaneously with a
(detuned) optical cavity \cite{nats, prls}. In these experiments,
the same coupling to the laser and cavity fields applies for ions
with a three-level $\Lambda$-type configuration [as displayed in
Fig.~\ref{fig:2}(b)]. For such a level configuration, the qubit is
associated with the states $\ket{1}$ and $\ket{a}$, and the W state
can be generated by moving the equally distanced trapped ions along
the trap. Similar as for the atomic chains above, the cavity-laser
mediated interaction between the ions is described by the effective
Hamiltonian (\ref{ham2}) and, therefore, requires the same analysis
as performed in Sects.~III.A - III.D in order to produce the W$_N$
states.

Finally, we like to mention our recent paper \cite{my} in which
another deterministic scheme for generation of the multipartite W
states has been proposed. In contrast to this work, the resonant
atom-cavity interaction regime has been exploited for $N$ flying
two-level (Rydberg) atoms which couple sequentially (one after
another) to the mode(s) of a high-finesse bimodal cavity.

\hspace{0.5cm}

\begin{acknowledgments}
This work was supported by the DFG under the project No. FR 1251/13.
\end{acknowledgments}

\end{document}